\documentclass[10pt,compsoc]{IEEEtran}
\usepackage{times}
\usepackage{amsmath}
\usepackage{siunitx}
\usepackage{graphicx}
\usepackage{mdwtab}
\usepackage{booktabs}
\usepackage{multirow}
\usepackage{multicol}
\usepackage{xcolor}
\usepackage{subcaption}
\usepackage{amssymb}
\usepackage{hyperref}
\usepackage{cite}
\usepackage{float}
\usepackage[justification=centering]{caption}

\usepackage[normalem]{ulem}
\usepackage[ruled,noresetcount,noend]{algorithm2e}
\usepackage{algpseudocode}
\usepackage{xcolor}
\definecolor{darkgreen}{rgb}{0,.4,0}
\definecolor{darkcyan}{rgb}{0,.4,.4}
\newcommand{\REMOVE}[1]%
          {{\color{red}\sout{#1}}}

\newcommand{\COMMENT}[1]%
          {{\color{darkgreen}\textbf{{PG:}} {#1}}}
          
\begin{document}
\title{Polyp-SAM++: Can A Text Guided SAM Perform Better for Polyp Segmentation?}
\author{Risab~Biswas
\thanks{R. Biswas is with Optiks Innovations Pvt. Ltd (P360), Mumbai, Maharashtra 400066, India. (e-mail: risab.biswas@p360.com).}}
%%%%%%%%%%%%%%%%%%%%%%%%%%%%%%%%%%%%
\markboth{Preprint Submitted to Arxiv, 2023}%
{Shell \MakeLowercase{\textit{et al.}}: Bare Demo of IEEEtran.cls for IEEE Journals}
% abstract and (if needed) index terms
\IEEEtitleabstractindextext{%
    \begin{abstract}
    Meta recently released SAM (Segment Anything Model) which is a general-purpose segmentation model. SAM has shown promising results in a wide variety of segmentation tasks including medical image segmentation. In the field of medical image segmentation, polyp segmentation holds a position of high importance, thus creating a model which is robust and precise is quite challenging. Polyp segmentation is a fundamental task to ensure better diagnosis and cure of colorectal cancer. As such in this study, we will see how \textbf{Polyp-SAM++}, a text prompt-aided SAM, can better utilize a SAM using text prompting for robust and more precise polyp segmentation. We will evaluate the performance of a text-guided SAM on the polyp segmentation task on benchmark datasets. We will also compare the results of text-guided SAM vs unprompted SAM. With this study, we hope to advance the field of polyp segmentation and inspire more, intriguing research. The code and other details will be made publically available soon at \url{https://github.com/RisabBiswas/Polyp-SAM++}.
    \end{abstract}
\begin{IEEEkeywords}
Medical Image Segmentation, Polyp Segmentation, Segment Anything Model
\end{IEEEkeywords}
}

% make the title area
\maketitle
\IEEEdisplaynontitleabstractindextext
\IEEEraisesectionheading{\section{Introduction}\label{sec:Intro}}
\IEEEPARstart{T}{ransformer}-based vision models like ViT~\cite{DBLP:journals/corr/abs-2010-11929} have been in the field for quite some time now and have performed well in tasks like medical image segmentation, image enhancement, object detection, classification, etc. In recent years, large language models like Chatgpt~\cite{NEURIPS2020_1457c0d6} and GPT-4~\cite{OpenAI2023GPT4TR} like models have gained quite popularity. This is mostly due to the fact that they are general-purpose and can be adapted easily for a wide variety of tasks. This is mostly due to the fact that they are trained on a large amount of data. LLMs provide the advantage of cross-modal learning and utilizing extensive data but may not be optimally used for image segmentation tasks. For such tasks, CNN and ViT-based models are still a preferred choice by researchers. 
Meta released the Segment Anything Model (SAM)~\cite{kirillov2023segment}, a universal segmentation model that utilizes the capabilities of "prompting" for image segmentation. SAM is trained on the Segment Anything 1-Billion mask dataset (SA-1B), which is also publically released by Meta. Prompts could be points on the image, a bounding box, text, or, more generally, any information suggesting what to segment in an image. SAM's main goal is to decrease the demand for task-specific modeling knowledge, training computing power, and customized data annotation for image segmentation problems. So far, SAM has gained promising results in segmentation tasks and more and more application areas are having implementation and experimental analysis of SAM. Tao~\textit{et al.}~\cite{yu2023inpaint} proposed a SAM-based image inpainting model, for removing/replacing/filling any object in an input image. SAM incorporated with stable diffusion also opened doors for more interesting projects with applications in image inpainting. SAMCOD~\cite{tang2023sam}, proposed by Tang \textit{et al.} evaluated the performance of SAM for the Camouflaged object detection (COD) task. Works proposed by Yao~\textit{et al.}~\cite{yao2023matte} and Li~\textit{et al.}~\cite{li2023matting} are more focused on using SAM for image matting tasks. 
% Ji~\textit{et al.} analyzed and discussed the benefits and challenges of SAM. SAM has also found its application in the field of robotics.

SAM has also found promising applications in the field of medical image segmentation. For example, Deng~\textit{et al.}~\cite{deng2023segment} proposed a detailed assessment of SAM on the zero-shot digital pathological segmentation task. Ma~\textit{et al.}were among the ones to propose the foundational models of using SAM for a universal medical image segmentation model, called MedSAM~\cite{ma2023segment}. They utilized various prompting to generate their results. MedSAM opened doors for further investigation of SAM's ability to perform well in more specific segmentation tasks.

\section{Related Works Based on SAM}
\begin{figure*}[ht]
    \centering
    \includegraphics[clip=true,trim = 00 00 00 00, width=0.99\textwidth]{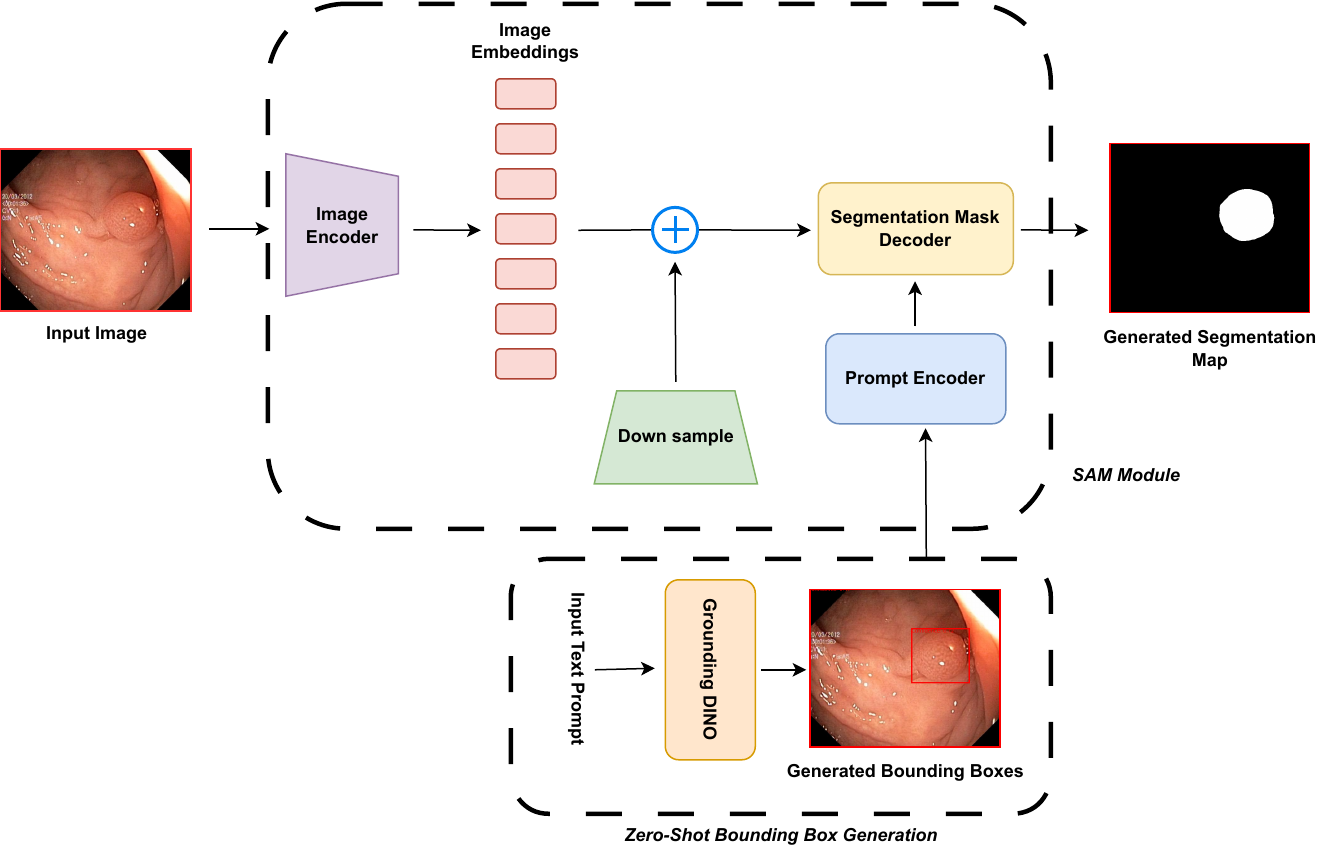} %TransDocUNet
    \caption{\centering \textbf{Overview of the Polyp-SAM++ Architecture}}
    \label{fig:Polyp-SAM_Arch}
\end{figure*}

Specifically for polyp segmentation, works done by Zhou~\textit{et al.}~\cite{zhou2023sam} and Chen~\textit{et al.}~\cite{chen2023sam} present promising results using an unprompted SAM and a domain-adapted SAM respectively. Additionally, Polyp-SAM~\cite{li2023polypsam} used SAM for the same task. Roy~\textit{et al.}~\cite{roy2023sammd} evaluated the zero-shot capabilities of SAM on the organ segmentation task. Inspired by these works on medical imaging, we performed an experiment to see how well a text-guided SAM (\textbf{Polyp-SAM++}) can segment polyps in colonoscopy images under various difficult circumstances. Particularly we provided a definition of a polyp with details about shape, color, position, etc. Our approach is very simple and robust. This report mainly evaluates the effectiveness of a text-guided SAM in segmenting polyps. Since Meta has not officially released the text prompting capabilities in SAM, we utilized Language Segment-Anything, which uses grounded-DINO to generate the bounding box based on the text prompt, which is eventually fed to the SAM for final segmentation mask generation. Fig.~\ref{fig:Polyp-SAM_Arch}, illustrates graphically how the Polyp-SAM++ model works. 
\section{Experiments and Results}
\label{sec:Results}
In this report, we compared the outcomes of \textbf{Polyp-SAM++} both quantitatively and qualitatively, with other state-of-the-art models on the polyp segmentation task as well as with previous outcomes when using SAM for polyp segmentation. In this section, we will discuss briefly the evaluation metrics, datasets, methods used for comparison, and implementation details of Polyp-SAM++. 
%%%%%%%%%%%%%%%%%%%%%%%%%%%%%%%%%%%%%%%%%%%%%%%%%
%%%%%%%%%%%%%%%%%%%%%%%%%%%%%%%%%%%%%%%%%%%%%%%%%
\subsection{Datasets}
\label{subsec:Dataset}
To evaluate the outcomes of \textbf{Polyp-SAM++}, we performed both qualitative and quantitative experiments on four benchmark colonoscopy datasets:
\begin{itemize}
    \item \textbf{Kvasir-SEG}~\cite{jha2020kvasir}: This dataset contains 1000 pairs of images from inside the gastrointestinal tract, during colonoscopy of patients, where the polyp is present along with the ground truth segmentation mask. The dataset has images with varied resolutions. The dataset is collected by Vestre Viken Health Trust, Norway. 
    \item \textbf{CVC-300}~\cite{vazquez2017benchmark}: This dataset contains 60 images containing polyp. The dataset contains images with a resolution of $500 \times 574$.  
    \item \textbf{CVC-ClinicDB}~\cite{bernal2015wm-dova}: This dataset contains 612 images collected from various colonoscopy videos with a resolution of $288 \times 384$.
\end{itemize}
% Among all the datasets used for our assessments, we found the Kvasir SEG dataset to be the most robust and high-quality dataset. We ran Polyp-SAM++ with the same text prompt for all the images. 

\subsection{Evaluation Metrics}
\label{subsec:Metrics}
In our experiment, we used the three most commonly used metrics for image segmentation to evaluate the effectiveness of Polyp-SAM++ and other methods. We used the Mean dice score($\rm mDice$), Mean intersection over union ($\rm mIoU$), and F-measure ($\rm Fm$)~\cite{5206596}. 

\subsection{Implementation Details}
\label{subsec:ImplDetails}
\begin{figure*}[ht!]
\centering
\begin{tabular}{ccccc}
{\includegraphics[width = 0.45\columnwidth]{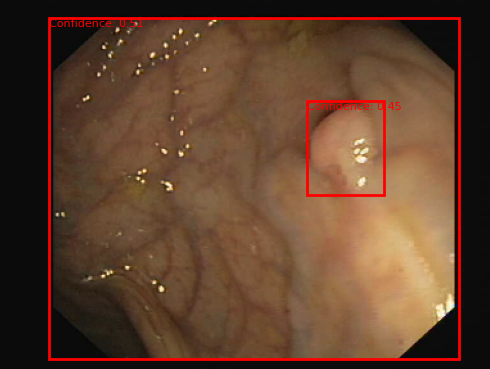}} &
{\includegraphics[width=0.45\columnwidth]{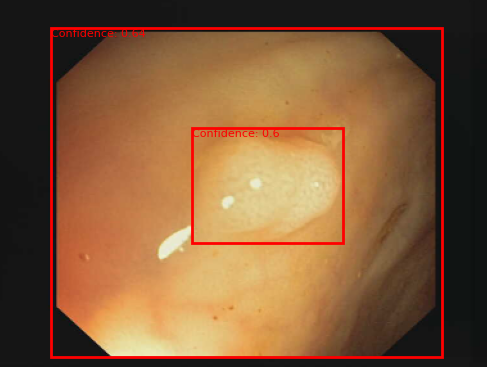}}&
{\includegraphics[width=0.45\columnwidth]{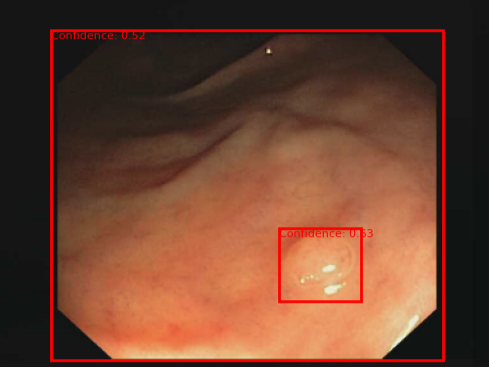}}&
{\includegraphics[width = 0.45\columnwidth]{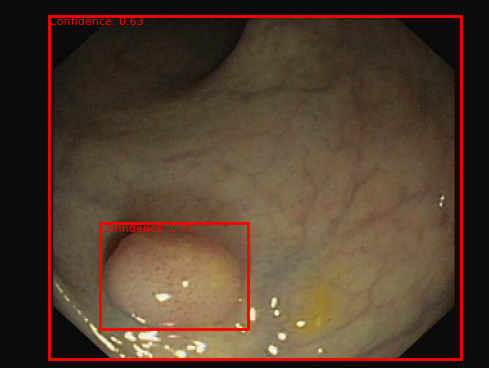}} & 
\end{tabular}
\caption{\textbf{Bounding Box created based on the Text-Prompt by GroundingDINO.}}
\label{fig:BoundingBox}
\end{figure*}

We begin our evaluation by handcrafting a text prompt that acts as a guide to the segment anything model. In our experiment for polyp segmentation, we are only concerned about generating the segmentation map of the polyp in the colonoscopy images. As discussed in the earlier sections, SAM is capable of handling a variety of prompts for generating specific segmentation maps in an image. One of them is by utilizing a bounding box around the objects of concern. Based on this idea, we are using our text-guided evaluation. The text prompt is crafted in a way that it defines a polyp which acts as a guide to the segmentation model. The text prompt creates a bounding box using GroundingDINO~\cite{liu2023grounding} which is a zero-shot approach. Fig.~\ref{fig:BoundingBox} illustrates the resultant bounding boxes. These bounding box helps the SAM model to generate a precise segmentation map of the object within the bounding box. During our evaluation, we noticed that when using SAM with text prompts, there were occurrences where it generated multiple bounding boxes in an image, resulting in more than one segmentation mask. In such situations, we devised a strategy to select the best mask based on the corresponding ground truth. Since we are not changing the text prompt at all throughout our experiment, it makes the overall evaluation more faster and robust. Though, we agree that in some cases where \textbf{Polyp-SAM++} did not generate good results or completely failed to produce a segmentation mask, input prompt engineering might overcome the shortcomings. We will provide some good examples of the failed scenarios as well in the later sections.

\textit{\textbf{Example Text Prompt} = "A polyp is an anomalous oval-shaped small bump-like structure, a relatively small growth or mass that develops on the inner lining of the colon or other organs. Multiple polyps may exist in one image."}

% %%%%%%%%%%%%%%%%%%%%%%%%%%%%%%%%%%%%%%%%%%%%%%%%%
% \begin{figure*}[ht!]
%     \centering
%     \includegraphics[clip=true, width=1.00\textwidth]{Results/BB.pdf} %TransDocUNet
%     \caption{\centering \textbf{Segmentation Results on Kvasir-SEG, CVC-300 and CVC-ClinicDB Datasets using Polyp-SAM++.}}
%     \label{fig:Polyp-SAM_Arch}
% \end{figure*}
% %%%%%%%%%%%%%%%%%%%%%%%%%%%%%%%%%%%%%%%%%%%%%%%%%

\begin{table*}[ht!]
\caption{\centering{\color{black}{\textbf{Quantitative Evaluation on the Benchmark Datasets for Polyp Segmentation.}}}}
\resizebox{\textwidth}{!}{
\centering
\begin{tabular}{l||lll|lll|lll} 
\midrule
\multirow{2}{*}{\textbf{Methods}} & \multicolumn{3}{c|}{\textbf{CVC-ClinicDB~\cite{bernal2015wm-dova}}} & \multicolumn{3}{c|}{\textbf{Kvasir-SEG~\cite{jha2020kvasir}}} & \multicolumn{3}{c}{\textbf{CVC-300~\cite{vazquez2017benchmark}}}  \\ 
\cline{2-10}

                         & $\rm mDice$$\uparrow$ & $\rm mIoU$$\uparrow$ & $\rm Fm$$\uparrow$   & $\rm mDice$$\uparrow$ & $\rm mIoU$$\uparrow$ & $\rm Fm$$\uparrow$ & $\rm mDice$$\uparrow$ & $\rm mIoU$$\uparrow$ & $\rm Fm$$\uparrow$ \\ 
\midrule
UNet~\cite{ronneberger2015unet}                 &0.82        &0.75       &0.81        &0.81        &0.746       &0.79        &0.71        &0.62       &0.68         \\
UNet++~\cite{zhou2018unet}               &0.79        &0.72       &0.78        &0.82        &0.74       &0.80        &0.70        &0.62       &0.68     \\
SFA~\cite{10.1007/978-3-030-32239-7_34}                  &0.70        &0.60       &0.64        & 0.72       &0.61       &0.67        &0.46        &0.32       &0.34\\
PraNet~\cite{fan2020pranet}               &0.89        &0.84       &0.89        &0.89        &0.84       &0.88        &0.87        &0.79       &0.84  \\
ACSNet~\cite{zhang2023adaptive}                 &0.88        &0.82       &0.87        &0.89        &0.83       &0.88        &0.86        &0.78       &0.82             \\
MSEG~\cite{huang2021hardnetmseg}               &0.90        &0.86       &0.90        &0.89        &0.83       &0.88        &0.87        &0.80       &0.85        \\
DCRNet~\cite{yin2022duplex}                  &0.89        &0.84       &0.89        &0.88        &0.82       &0.86        &0.85        &0.78       &0.83           \\
EU-Net~\cite{patel2021enhanced}               &0.90        &0.84       &0.89        &0.90        &0.85       &0.89        &0.83        &0.76       &0.80             \\
SANet~\cite{wei2021shallow}                 &0.91        &0.85       &0.90        &0.90        &0.84       &0.89        &0.88        &0.81       &0.80          \\
MSNet~\cite{Zhao2021AutomaticPS}               &0.91        &0.86       &0.91        &0.90        &0.84       &0.89        &0.86        &0.79       &0.84           \\
C2FNet~\cite{sun2021contextaware}                  &0.91        &0.87       &0.90        &0.88        &0.83       &0.87        &0.87        &0.80       &0.92        \\
LDNet~\cite{zhang2023lesionaware}               &0.88        &0.82       &0.87        &0.88        &0.82       &0.86        &0.86        &0.79       &0.84          \\
FAPNet~\cite{Zhou_2022}                 &0.92        &0.87       &0.91        &0.90        &0.84       &0.89        &0.89        &0.82       &0.87             \\
CFA-Net~\cite{ZHOU2023109555}               &0.93        &0.88       &0.92        &0.91        &0.86       &0.90        &0.89        &0.82       &0.87          \\ 
Polyp-PVT~\cite{dong2023polyppvt}                  &\textbf{0.94}       &\textbf{0.90}       &\textbf{0.95}        &\underline{0.91}        &0.86       &0.91        &0.90        &\textbf{0.93}       &\textbf{0.88}             \\
HSNet~\cite{ZHANG2022106173}               &\underline{0.93}        &\underline{0.88}       &\underline{0.93}        &\textbf{0.92}        &\textbf{0.87}       &\underline{0.91}        &0.90        &0.83       &\textbf{0.88}          \\
Polyp-SAM~\cite{li2023polypsam}               &0.92        &0.87       & -        &0.90        &0.86       & -        &\textbf{0.92}        & \underline{0.88}      & -         \\
\toprule
SAM-H~\cite{zhou2023sam}               &0.54        &0.50       &0.54        &0.77        &0.70       &0.76        &0.65        &0.60       &0.65          \\
SAM-L~\cite{zhou2023sam}               &0.57        &0.52       &0.56        &0.78        &0.71       &0.77        &0.72        &0.67       &0.72          \\
\toprule
\toprule
\textbf{Polyp-SAM++}               &{0.91}        &{0.86}        &{0.91}         &{0.90}        &\underline{{0.86}}       &\textbf{{0.92}}        & {0.73}        &{0.69}        &{0.73} \\
\bottomrule
\label{Tab:Table1}
\end{tabular}}
\end{table*}

%----------16 2017--------------

%%%%%%%%%%%%%%%%%%%%%%%%%%%%%%%%%%%%%%%%%%%%%%%%%
\begin{figure*}[ht!]
    \centering
    \includegraphics[clip=true, trim = 00 00 00 00, width=1.00\textwidth]{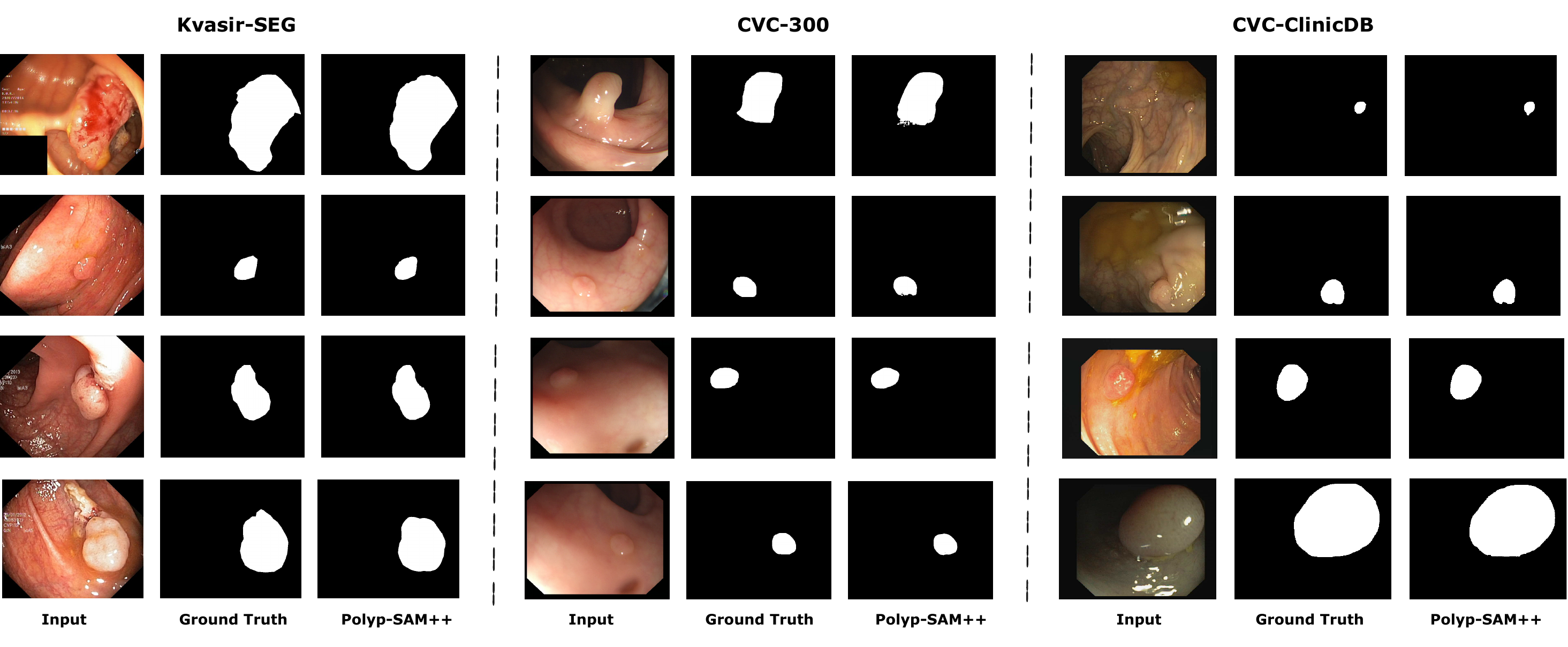} %TransDocUNet
    \caption{\centering \textbf{Qualitative Assessment of Segmentation Outcomes on Kvasir-SEG, CVC-300 and CVC-ClinicDB Datasets using Polyp-SAM++.}}
    \label{fig:Polyp-Outcome}
\end{figure*}
%%%%%%%%%%%%%%%%%%%%%%%%%%%%%%%%%%%%%%%%%%%%%%%%%
%%%%%%%%%%%%%%%%%%%%%%%%%%%%%%%%%%%%%%%%%%%%%%%%%
\begin{figure*}[ht!]
    \centering
    \includegraphics[clip=true, trim = 00 00 00 00, width=1.00\textwidth]{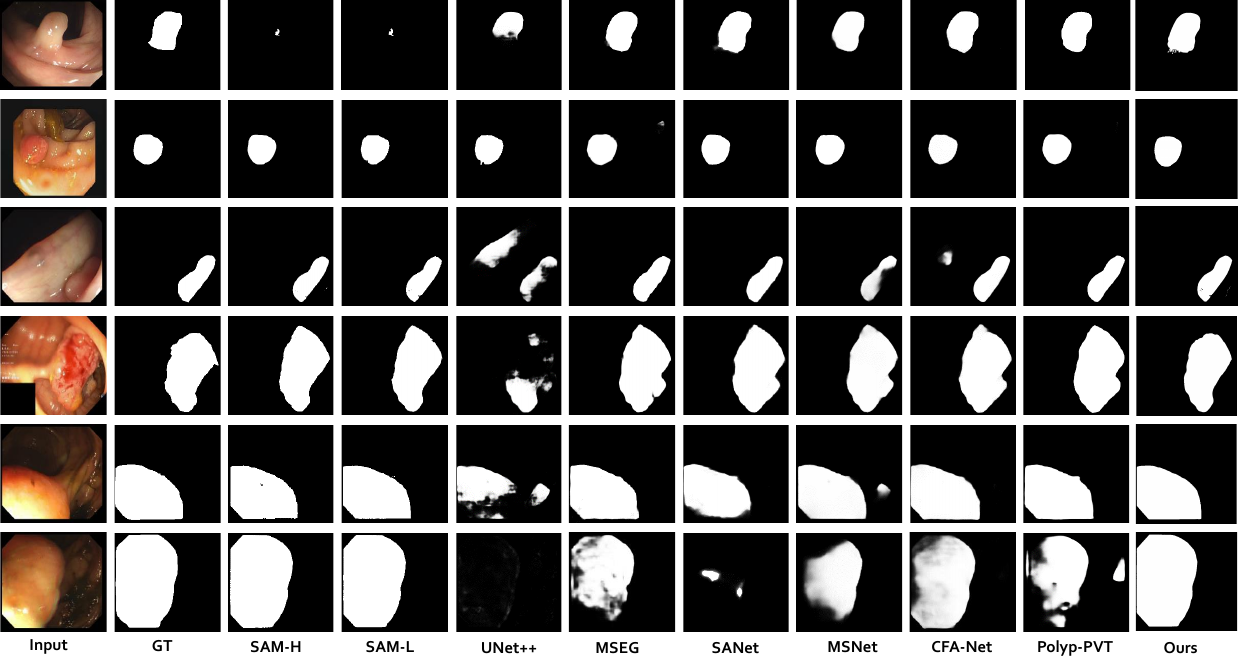} %TransDocUNet
    \caption{\centering \textbf{Qualitative Analysis on the Segmentation Outcomes and Comparison using Polyp-SAM++.}}
    \label{fig:Polyp-SAM++ Results}
\end{figure*}
%%%%%%%%%%%%%%%%%%%%%%%%%%%%%%%%%%%%%%%%%%%%%%%%%
%----------2011-----------
\begin{figure}[ht!]
\centering
\resizebox{5cm}{!}{
\begin{tabular}{cccc}
\includegraphics[width=0.5\columnwidth]{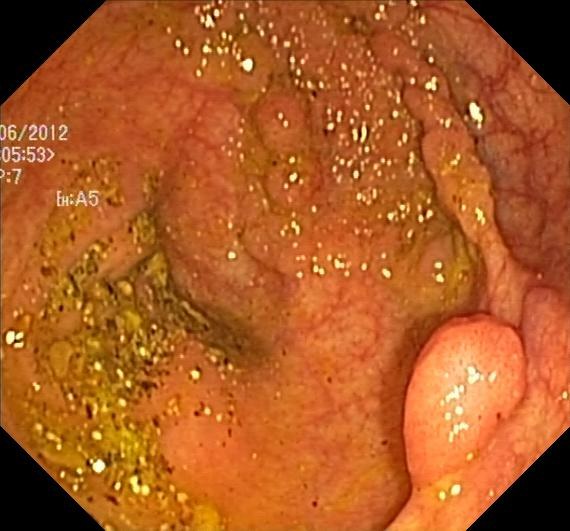} &
\includegraphics[width=0.5\columnwidth]{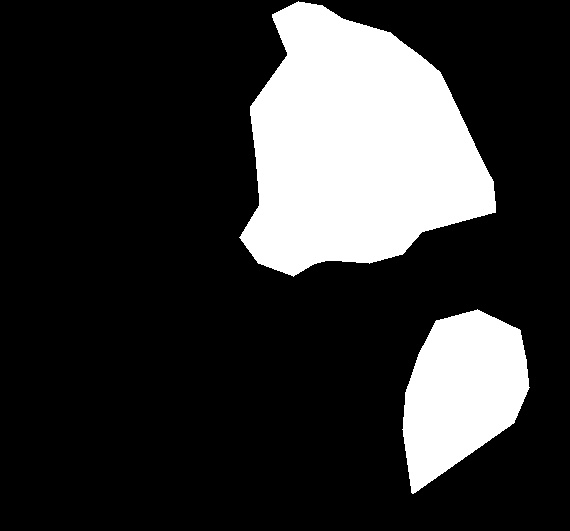} &
\includegraphics[width=0.5\columnwidth]{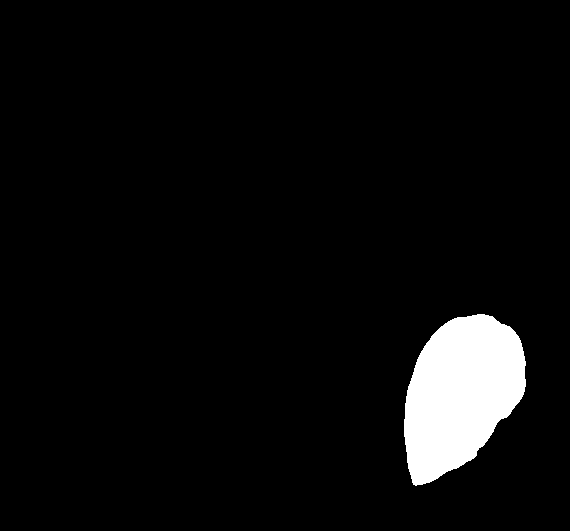} & \\
\includegraphics[width=0.5\columnwidth]{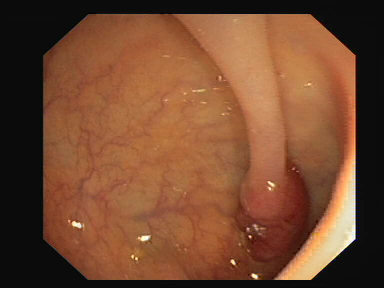} &
\includegraphics[width=0.5\columnwidth]{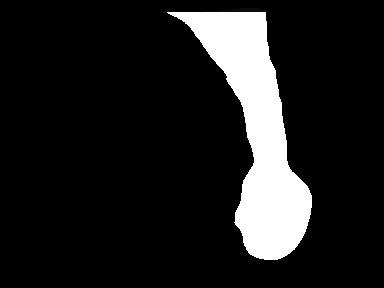} &
\includegraphics[width=0.5\columnwidth]{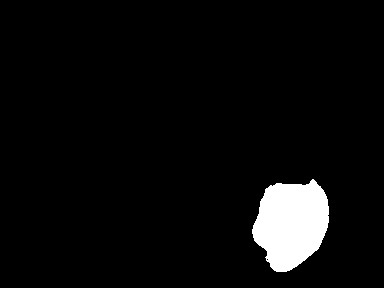} & \\
\includegraphics[width=0.5\columnwidth]{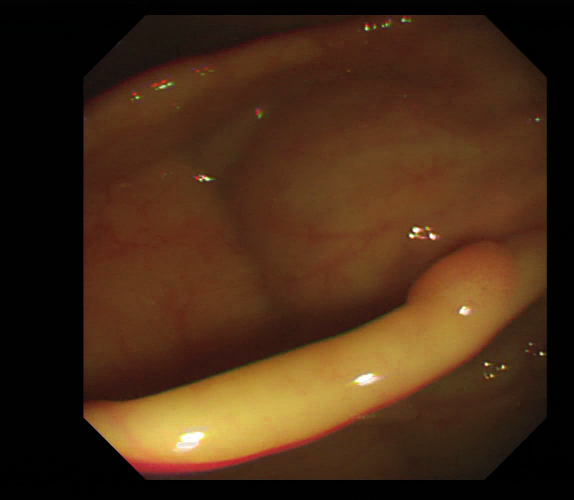} &
\includegraphics[width=0.5\columnwidth]{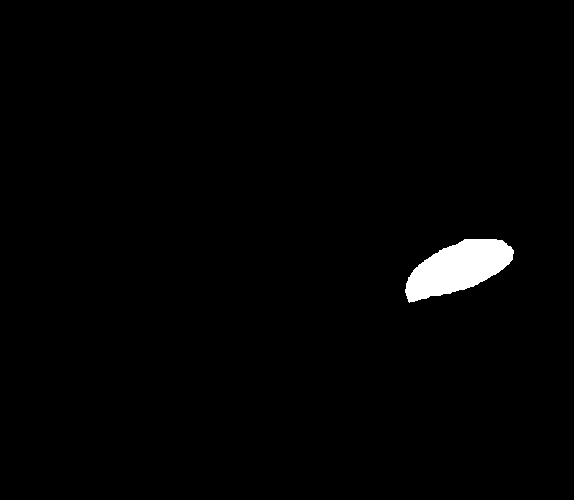} &
\includegraphics[width=0.5\columnwidth]{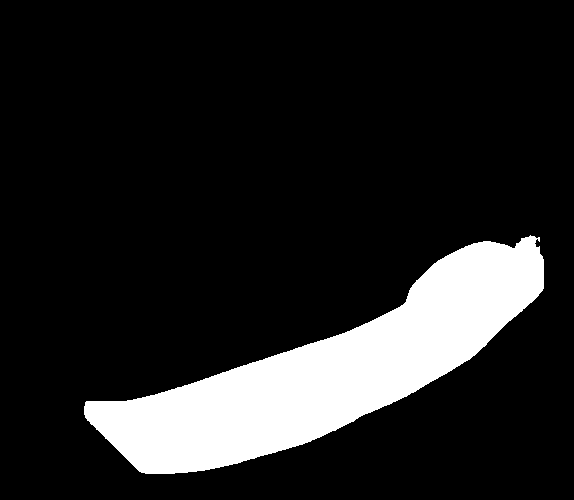} & \\
\end{tabular}}
\caption{{\textbf{Cases where Polyp-SAM++ failed on generating accurate segmentation masks.}}}
\label{fig:FailureCases}
\end{figure}

For the quantitative evaluation, we have randomly chosen 100 images from the Kvasir-SEG dataset, and 60 random images from the CVC-ClinicDB. For evaluation, all the images from CVC-300 are used. 
\subsection{Methods for Comparison}
\label{subsec:Methods}
For our study, we compared the results produced by \textbf{Polyp-SAM++} and compared them with the outcomes of various CNN, ViT, and even other recent SAM-based techniques for polyp segmentation. The CNN based models include - UNet~\cite{ronneberger2015unet}, UNet++~\cite{zhou2018unet}, SFA~\cite{10.1007/978-3-030-32239-7_34}, PraNet~\cite{fan2020pranet}, ACSNet~\cite{zhang2023adaptive}, MSEG~\cite{huang2021hardnetmseg}, DCRNet~\cite{yin2022duplex}, EU-Net~\cite{patel2021enhanced}, SANet~\cite{wei2021shallow}, MSNet~\cite{Zhao2021AutomaticPS}, and CFA-Net~\cite{ZHOU2023109555} . For the Transformer based models, we assessed Polyp-PVT~\cite{dong2023polyppvt} and 
HSNet~\cite{ZHANG2022106173}. And finally, we assessed the work done by Tao \textit{et al.}~\cite{zhou2023sam} and Polyp-SAM~\cite{li2023polypsam}, where they evaluated the performance of SAM for polyp segmentation. 

\subsection{Quantitative Results and Comparison}
In this section, we will look into how Polyp-SAM++ performs in polyp segmentation tasks from a quantitative evaluation perspective. We compared the outcomes with the methods mentioned in~\ref{subsec:Methods} on the three datasets mentioned in the ~\ref{subsec:Dataset} based on the metrics discussed in~\ref{subsec:Metrics}. The results of the model and a comparative analysis are presented in Table~\ref{Tab:Table1}. The best results are highlighted in bold and the second best is underlined. The Polyp-SAM model which is a SAM-based architecture fine-tuned specifically for polyp segmentation and SAM-H/SAM-L are also compared with the Polyp-SAM++ results. We can evidently see that using simple text guidance to generate the bounding box helped improve the SAM model to better understand what to segment, which improved the overall segmentation performance. The text guidance is helping in the localization while the SAM is performing the segmentation. Based on the analysis we can understand that the shortcoming of using a model like SAM is on the localization objective because once the localization is achieved, SAM is effectively able to segment that. The text guidance is not only helping in the localization of a polyp in the input image but also aids in generating a better segmentation map.    

\subsection{Qualitative Results and Comparison}
In this section, we will look into how Polyp-SAM++ performs in polyp segmentation tasks from a qualitative evaluation perspective. The Fig.~\ref{fig:Polyp-SAM++ Results} shows the results of the text-guided SAM on polyp segmentation on the three chosen benchmark datasets. We can see how the model effectively understands the text prompt and segments the region of interest. The Polyp-SAM++ performs segmentation well in diverse settings without making changes in the prompt, this shows the robustness of the GroundingDINO and the SAM methods. Next we compared the outcomes with the results of the other state-of-the art methods mentioned in~\ref{subsec:Methods}. Based on the outcomes illustrated in Fig.~\ref{fig:Polyp-Outcome}, it is evident that the text guidance is helping the segmentation model overcome the challenges faced by the SAM outcomes without any prompting about the object of interest. And overall the model is quite competitive to the current polyp segmentation models. However, even with the text guidance SAM fails to segment the polyps in certain cases. A few examples are shown in Fig.~\ref{fig:FailureCases}. We will discuss how can we overcome this challenge in the next section~\ref{section:future_work}.  
% Provide insights here 

\section{Future Work}
\label{section:future_work}
Even though Polyp-SAM++ has shown impressive performance as a strong competitor in the field of cutting-edge models for polyp segmentation, there are still exciting opportunities to make it even better. In this part, we'll explore potential ways to enhance and fine-tune the Polyp-SAM++ model for even more remarkable results.

\begin{itemize}
\item \textbf{Evaluation on Other Datasets:} It is essential to subject Polyp-SAM++ to evaluation using the \textit{CVC-ColonDB}~\cite{tajbakhsh2015automated} and \textit{ETIS}~\cite{silva2014toward} datasets, both of which serve as benchmark datasets in this domain. This evaluation process will provide crucial insights into the model's performance and its ability to handle these established and recognized datasets effectively.
\item \textbf{Zero-shot Bounding Box Generator Enhancement:} A significant opportunity for enhancement involves improving the zero-shot bounding box generator within Polyp-SAM++. By refining this component using a broader collection of annotated polyp images, the model's skill in generating precise bounding boxes within colonoscopic images can be substantially improved. This advancement would establish a more resilient initial localization of potential polyps, subsequently resulting in heightened accuracy during the subsequent segmentation process. 

\item \textbf{Fine-Tuning SAM:} Similar to the approach adopted for the original Polyp-SAM model, the Polyp-SAM++ model can benefit from fine-tuning its underlying SAM model. This process involves training the model on a specialized dataset of annotated polyps to better adapt its capabilities to the unique characteristics of polyp segmentation tasks. This fine-tuning can enable the model to capture the nuances of polyp appearances and further enhance its segmentation accuracy.

\item \textbf{Prompt Engineering for Human-Like Supervision: }Exploring prompt engineering offers an intriguing way for refining the Polyp-SAM++ model. By crafting specialized prompts for different datasets or scenarios, the model's performance can be guided more effectively. This approach essentially introduces a form of human-like supervision to guide the model's decision-making, potentially improving its understanding of polyp-specific features and characteristics.

\item \textbf{Performance Evaluation with Meta's Text Prompts:} Once Meta officially releases their text prompting capability in SAM, an evaluation of the Polyp-SAM++ model's performance using these prompts can provide insights into the model's adaptability and versatility. This assessment can guide further refinements and adaptations to leverage the benefits of the Meta text prompts for polyp segmentation.
\end{itemize}

\section{Conclusion}
This report provides an investigation of the performance of SAM using a text-guided input prompt for the polyp segmentation task. We tested \textbf{Polyp-SAM++} on three standard polyp-segmentation datasets, and the outcomes presented a quantitative and qualitative evaluation and comparison respectively. Based on the results and our analysis, we see by taking a simple approach of incorporating text prompts to SAM can increase the overall performance significantly. We discussed that a general-purpose model like SAM can still be used for a specific task like polyp segmentation by providing SAM the relevant context, in this case, a bounding box to find the region of interest as the localization step in the overall task. We also show quantitatively and qualitatively how the text guidance improves upon the SAM without any prompting for polyp segmentation. We discussed the failed cases as well as the model and eventually discussed how we can further improve the model. We are hoping this paper will encourage more curiosity in the subject of polyp segmentation and that further research will be carried out using SAM to make progress in this field of study.

\section{Acknowledgment}
The author would like to extend a heartfelt appreciation to Luca Medeiros, the author and creator of the repository "lang-segment-anything" available at \url{https://github.com/luca-medeiros/lang-segment-anything}. The remarkable work done by the author has been instrumental in making our project possible and has greatly contributed to its success.
\bibliographystyle{IEEEtran}
\bibliography{refs}
\end{document}